\documentstyle[twocolumn]{jpsj}

\def\ggs{\buildrel\textstyle > \over {\hbox{\raise0.2ex\hbox{$\sim$}}}}
\def\lls{\buildrel\textstyle < \over {\hbox{\raise0.2ex\hbox{$\sim$}}}}
\def\gsim{\,\lower0.75ex\hbox{$\ggs$}\,}
\def\lsim{\,\lower0.75ex\hbox{$\lls$}\,}

\newcommand{\Js}{J_{\rm s}}

\newcommand{\ii}{{\rm i}}

\newcommand{\kb}{{\bf k}}
\newcommand{\qb}{{\bf q}}

\newcommand{\tk}{{t_{\kb}}}

\newcommand{\simg}{\stackrel{>}{_\sim}}
\newcommand{\siml}{\stackrel{<}{_\sim}}

\def\runtitle{Pseudogap Induced by Superconducting Fluctuation in the $d$-$p$ Model}
\def\runauthor{Akito {\sc Kobayashi} {\it et al.}}

\title
{Pseudogap Induced by \\ Superconducting Fluctuation in the $d$-$p$ Model}

\author
{
Akito {\sc Kobayashi}$^{1, 2}$, Atsushi {\sc Tsuruta}$^{1}$, 
Tamifusa {\sc Matsuura}$^1$ and Yoshihiro {\sc Kuroda}$^{1, 2}$, 
}

\inst
{
$^{1}$Department of Physics, Nagoya University, Nagoya 464-01 \\
$^{2}$CREST, Japan Science and Technology Corporation (JST) \\
}

\recdate{
\hspace*{20mm}
}

\abst{
Using the $d$-$p$ model, we demonstrate that the pseudogap, which is induced by the superconducting fluctuation, plays key roles in the determination of the phase diagram observed in high-$T_{\rm c}$ superconducting materials.
We take the pairing interaction mediated by the spin fluctuation and calculate the superconducting transition temperature $T_{\rm c}$, the NMR relaxation rate $1/T_1$ and the single-particle spectrum by treating both the superconducting fluctuation and spin fluctuation in a consistent fashion.
As temperature decreases, $1/T_1T$ increases at high temperatures, and it reaches a maximum followed by a sharp drop in the underdoped region, due to the evolution of the pseudogap in the single-particle spectrum.
The evolution is also consistent with those of ARPES experiments.
}

\kword{pseudogap, NMR relaxation rate, superconductivity, superconducting fluctuation, spin fluctuation, superexchange interaction, $d$-$p$ model, $1/N$-expansion, slave-boson technique} 

\begin{document}
\input epsf

\sloppy
\maketitle

%%%%%%%%%%%%%%%%%%%%%%%%%

%\section{Introduction}

%%%%%%%%%%%%%%%%%%%%%%%%%

In the high-$T_{\rm c}$ superconductors, various anomalous properties have been observed at higher temperatures than the superconducting transition temperature $T_{\rm c}$, especially in the underdoped region.
Key issues on these anomalous properties must be on the existence of the pseudogap which has been suggested based on various experiments.~\cite{Yasuoka,Rossat-Mignod,Takenaka,Nishikawa,Nakano,Sato,Loeser,Ding,Norman}
In-plane resistivities,\cite{Takenaka} Hall coefficients\cite{Nishikawa} and static uniform spin susceptibilities\cite{Nakano} show peculiar temperature dependences below a certain characteristic temperature $T_0$,\cite{Sato} where the pseudogap is believed to become appreciable.
Recently, photoemission experiments have revealed that the pseudogap is of the same size and has the same $\kb$-dependence as the superconducting gap.\cite{Loeser,Ding} 
More recently, the temperature evolution of the pseudogap has been observed by the angle-resolved photoemission spectroscopy (ARPES).\cite{Norman} 

There are several different views regarding the origin of the pseudogap.
For example, it may be interpreted as a spin-excitation gap in the singlet RVB states obtained in the $t$-$J$ model,\cite{Suzumura,Fukuyama} or it may be due to the Bose condensation of bound pairs\cite{Randeria,Koikegami} or due to superconducting fluctuations\cite{Emery,Doniach,Levin,Schafer}.

In the present study, we demonstrate, using the $d$-$p$ model, that the pseudogap induced by the superconducting fluctuation (SC fluctuation) plays key roles in the determination of the phase diagram observed in high-$T_{\rm c}$ superconducting materials.
%The superconducting fluctuation is due to the pair fluctuation of quasi-particles in the in-gap band, whose band width is proportional to hole-doping rate $\delta$ at the underdoped region.~\cite{Jichu,Ono2}
%The superconducting fluctuation is calculated in the in-gap band, whose band width is proportional to hole-doping rate $\delta$ at the underdoped region.~\cite{Jichu,Ono2}
The pairing interaction is derived from the antiferromagnetic spin fluctuation (AF fluctuation) induced by the superexchange interaction $J_{\rm s}$, while the AF fluctuation is suppressed by the pseudogap in the single-particle spectrum induced by the SC fluctuation.
%Here we should notice that the AF fluctuation is suppressed by the pseudogap, hence the pairing interaction is modified.
Thus, we must treat the effects of both the SC fluctuation and the AF fluctuation in a consistent fashion.
%By extending the self-consistent $t$-matrix approximation, we treat both the AF and SC fluctuations self-consistently and calculate the superconducting transition temperature $T_{\rm c}$, the single-particle spectra and NMR relaxation rate $1/T_1 T$.
Thus, we calculate the superconducting transition temperature $T_{\rm c}$, the NMR relaxation rate $1/T_1$ and the single-particle spectrum.
$T_{\rm c}$ has a maximum at the hole-doping rate $\delta \cong 0.11$.
At high temperatures, $1/T_1T$ increases as $T$ decreases, due to the development of the AF fluctuation.
We assign $T_{\rm 0}$ as a temperature at which the effects of the AF fluctuation begin to be appreciable.
Then this $T_{\rm 0}$ corresponds to that determined by experiments on the Hall coefficient $R_{\rm H}$.~\cite{Nishikawa}
As the temperature decreases further, $1/T_1T$ reaches a maximum at a certain temperature, denoted by $T_{\rm sg}$, for $\delta < 0.11$, followed by a sharp drop at $T \siml T_{\rm sg}$.
These features are consistent with the results of NMR experiments.~\cite{Yasuoka,Rossat-Mignod}
The plots of these three characteristic temperatures, $T_{\rm c}$, $T_{\rm 0}$ and $T_{\rm sg}$, as functions of $\delta$, account for essential features of the phase diagram of the high-$T_{\rm c}$ superconducting materials.
%We may argue that $T_{\rm 0}$ obtained in the present study corresponds to $T_{\rm 0}$ ~\cite{Nishikawa} because the enhancement of $R_{\rm H}$ has shown to be due to AF fluctuation~\cite{Kontani} as same as that of $1/T_1 T$.
%Hence the phase diagram qualitatively accounts for the anomalous metallic phase observed in high-$T_{\rm c}$ superconducting materials.~\cite{Sato}
%We note that the SC fluctuation becomes appreciable at $T \sim T_{\rm 0}$.
The single-particle spectral weights at the Fermi level decrease near the $[\pi ,0]$ and $[0,\pi ]$ directions as $T$ decreases at $T \siml T_{\rm 0}$, and eventually the Fermi surface survives only near the $[\pi ,\pi ]$ direction at $T \siml T_{\rm sg}$.
These features are consistent with the results of ARPES experiments.\cite{Norman}

First, we describe our model and formulation.
We take the simplest version of the $d$-$p$ model as
%The essential elements to be included are energies of a Cu-3$d$ orbital and a O-2$p$ orbital, transfer energy between a nearest neighbour pair of Cu-3$d$ and O-2$p$ orbitals and the intraatomic Coulomb repulsion between electrons within a single Cu-site, denoted as $\varepsilon_d, \varepsilon_p, t$, and $U$, respectively. 
%Furthermore, by taking account of the fact that the value of $U$ is much larger than the charge transfer gap $\Delta \equiv |\varepsilon_p - \varepsilon_d|$ in real systems, the simplest version of the  model is given by 
\begin{eqnarray} 
H &=& \varepsilon_p\sum_{m=1}^{N}\sum_{\kb_{m}}c_{\kb_m m}^+c_{\kb_m m} 
       +  \varepsilon_d\sum_{m=1}^{N}\sum_{i}d_{im}^+d_{im}
\nonumber \\       
&+& N_L^{-\frac12}\sum_{m=1}^{N}\sum_{\kb_m, i}\{t_i(\kb_m)
           c_{\kb_m m}^+d_{im}b_i^+ + h.c.\}, \label{hamiltonian}
\end{eqnarray}
which is treated within the physical subspace where local constraints
\begin{eqnarray} \label{local constraint}
  \hat{Q}_i &\equiv& \sum_{m=1}^{N}d_{im}^+d_{im} + b_i^+b_i = 1 \ \ ( i=1,2,\cdots,N_L ) \end{eqnarray}
hold.
In the above, $c_{\kb_m m},  d_{im}$ and  $b_i$ are annihilation operators for a $p$-hole, a $d$-hole and a slave boson, respectively, and $t_i(\kb) = t_{\kb} \exp(-i\kb\cdot{\bf R}_i)$, where $t_{\kb} \equiv  2t[1-\frac12 (\cos k_{x}a + \cos k_{y}a)]^\frac12$ and $a$ is the lattice constant.
The suffix $m$ represents the spin-orbital degeneracy $(m=1, 2, \cdots, N)$ introduced by dividing the phase space for $\kb$ into $N/2$ subspaces for $\kb_m$, keeping the total degrees of freedom for spins and orbitals unchanged as $\sum_{\sigma, \kb} = \sum_{m, \kb_m}$.\cite{Ono} 
For numerical calculations in the present study, we set $N=2$. 
The local constraints in eq. (\ref{local constraint}) are strictly held when we calculate an expectation value of a physical quantity.\cite{Coleman,Jin}

%In order to guarantee the local constraints in eq. (\ref{local constraint}) to hold strictly, we calculate an expectation value of a physical quantity $\hat{A}$ in the following manner,\cite{Coleman,Jin}
%\begin{eqnarray}
%&& \langle\hat{A}\rangle = \lim_{\{\lambda_i\}\to\infty}
%                           \langle\hat{A}\prod_i\hat{Q}_i\rangle_\lambda
%                           /\langle\prod_i\hat{Q}_i|\rangle_\lambda, 
%\\
%&& \langle\hat{A}\rangle_\lambda \equiv 
%               \mbox{Tr}[e^{-\beta H_{\lambda}}\hat{A}]/
%               \mbox{Tr}[e^{-\beta H_{\lambda}}], 
%\\
%&& H_{\lambda} \equiv H + \sum_i\lambda_i\hat{Q}_i.
%\end{eqnarray}
%In the present study, we strictly follow the above procedure. 

A set of self-consistent equations for single-particle Green's functions of the leading order in the $1/N$-expansion was solved to yield the $p$-hole Green's function given by~ \cite{Jichu,Ono2}
\begin{eqnarray} \label{GK} 
G(\kb, \ii \varepsilon_n) = \sum_{\gamma=\pm}\frac{A^{\gamma}_{\kb}}{\ii \varepsilon_n - E^{\gamma}_{\kb}}, 
\end{eqnarray}
with
\begin{eqnarray}
E^{\gamma}_{\kb}&\equiv&\frac12\left(\varepsilon_{p}+\omega_0+\gamma\sqrt{(\varepsilon_{p}-\omega_0)^2+4b \ \tk^2}\ \right), \label{EK} \\
A^{\gamma}_{\kb} &\equiv& \gamma \frac{E^\gamma_\kb - \omega_0}{E^+_\kb - E^-_\kb}, \label{AK}  
\end{eqnarray}
where $\omega_0$ and $b$ are the energy and the residue of the pole in the slave-boson Green's function.
%\begin{eqnarray}  
%& & -\frac{1}{\pi}{\rm Im}B (\varepsilon_d {+}\lambda_i{-}\omega{+}\ii0^+)
%= b\ \delta(\omega{-}\omega_0 ) \nonumber \\ 
%& & \hspace{3cm} +d\ \delta(\omega{-}\omega_d ) + C(\omega). \label{BI}
%\end{eqnarray}
%Both $\omega_0 \ ( \ >0)$ and $\omega_d \ ( \ <\varepsilon_d <0)$ are solutions of the equation for $\omega$ as  
%\begin{eqnarray}
%\omega-\varepsilon_d = \frac1{N_L} \sum_{\kb \sigma \gamma}\frac{t_\kb^2 \ A^\gamma_\kb \ f(E^\gamma_\kb)}{\omega-E^\gamma_\kb }, \label{E0}  
%\end{eqnarray}
%while the residues, $b$ and $d$, are  determined by 
%\begin{eqnarray}
%& & \frac1b = 1 + \frac1{N_L} \sum_{\kb \sigma \gamma}\frac{t_\kb^2 \ A^\gamma_\kb \ f(E^\gamma_\kb)}{(\omega_0-E^\gamma_\kb)^2}, \label{b} \\
%& & \frac1d = 1 + \frac1{N_L} \sum_{\kb \sigma \gamma}\frac{t_\kb^2 \ A^\gamma_\kb \ f(E^\gamma_\kb)}{(\omega_d-E^\gamma_\kb)^2}, \label{d} 
%\end{eqnarray} 
%The third term in eq. (\ref{BI}), $C(\omega)$, represents continuous spectra, which is finite for ${\rm Min}\{E^{-}_{\kb} \}\leq\omega\leq 0$. 
The chemical potential $\mu$ is determined by 
\begin{eqnarray}
n=1+\delta = \frac1{N_L} \sum_{\kb \sigma \gamma} \ f(E^\gamma_\kb),  \label{np} 
\end{eqnarray} 
where $n$ and $\delta$ are the total hole number and the doped-hole number per unit cell, respectively, and $f(E) = [ \ \exp\{E/k_{\rm B}T\} + 1 \ ]^{-1}$. 
Here, it is noted that the solution $E^{-}_{\kb}$ denotes the in-gap state emerging inside the charge transfer gap, $\Delta \equiv \varepsilon_p - \varepsilon_d$, upon doping carriers to a Mott insulator.
The band width of the in-gap states is proportional to $\delta$ in the underdoped reigon.
The effects of the interactions among the in-gap states are of higher-order terms in the $1/N$-expansion.~\cite{Miura,Hirashima,Azami2,Kobayashi,Kobayashi2,Azami3,Kobayashi3} 
Recently, it has been shown that quasi-particle interactions via the superexchange interaction play dominant roles in underdoped systems.~\cite{Fukagawa}

%%%%%%%%%%%%%%%%%%%%%%%%%%%%%%%%%%%%%%%%%%%%%%

%\section{Effects of Superconducting Fluctuation}

%%%%%%%%%%%%%%%%%%%%%%%%%%%%%%%%%%%%%%%%%%%%%%

Now, we derive coupled equations to treat the effects of the SC fluctuation and the the AF fluctuation in a consistent fashion (termed the extended self-consistent $t$-matrix approximation).
The pairing interaction mediated by the AF fluctuation via the superexchange interaction $J_{\rm s}$ in RPA is given by~\cite{KobayashiNSR}
\begin{equation}
V({\bf q}) \cong \frac{J_{\rm s} ({\bf q})}{1-J_{\rm s} ({\bf q}) \chi_{\rm s}^{(0)} ({\bf q},0)},
\end{equation}
with $J_{\rm s} ({\bf q})=-J_{\rm s} (\cos (q_x )+\cos (q_y ))$,
\begin{eqnarray}
\chi_{\rm s}^{(0)} ({\bf q},\omega +{\rm i}0^+ ) \cong \frac{b^2}{N_L}\sum_{\bf k} \frac{t_{{\bf k}+{\bf q}}^2 t_{\bf k}^2 }{(E_{{\bf k}+{\bf q}}^- -\omega_0 )^2 (E_{\bf k}^- -\omega_0 )^2 } \nonumber \\
\times \frac{1}{\pi}\int_{-\infty}^{\infty}{\rm d}x f(x) \times [ {\rm Im} G ({\bf k}+{\bf q},x+{\rm i}0^+ ) G ({\bf k},-\omega +x-{\rm  i}0^+ ) \nonumber \\
 + G ({\bf k}+{\bf q},\omega +x+{\rm i}0^+ ) {\rm Im} G ({\bf k},x+{\rm i}0^+ )].
\end{eqnarray}
It was shown that a component with the $d_{x^2 -y^2}$ symmetry among various components of the spin-fluctuation-mediated interaction contributes dominantly to the pairing interaction.\cite{Azami3}  
Therefore, in the present study, we take only the component with the $d_{x^2 -y^2}$ symmetry of the pairing interaction as
\begin{equation}
v_d =\frac{1}{N_L^2} \sum_{\bf k} \sum_{\bf k^\prime} \psi_d ({\bf k}) V({\bf k}-{\bf k^\prime}) \psi_d ({\bf k^\prime}),
\end{equation}
with $\psi_d ({\bf k})=\cos (k_x a)-\cos (k_y a)$.

The pairing susceptibility $\chi({\bf q}, \omega )$, which is the irreducible part of the $t$-matrix, is given by
\begin{eqnarray}
\chi ({\bf q},\omega +{\rm i}0^+ )=-\frac{1}{N_L}\sum_{\bf k}\psi_d ({\bf k})^2 \frac{b^2 t_{\bf k}^2 t_{{\bf q}-{\bf k}}^2}{((\pi T)^2 +\omega_0^2 )^2} \frac{1}{\pi} \int_{-\infty}^\infty {\rm d}x  \nonumber \\
\times [ f(x){\rm Im} G ({\bf k},x+{\rm i}0^+ ) G ({\bf q}-{\bf k},\omega -x+{\rm i}0^+ ) \nonumber \\
 -f(-x) G ({\bf k},\omega -x+{\rm i}0^+ ) {\rm Im} G ({\bf q}-{\bf k},x+{\rm i}0^+ )].
\end{eqnarray}
A self-energy correction due to the SC fluctuation in the $t$-matrix approximation is given by
\begin{eqnarray}
\Sigma ({\bf k},\omega +{\rm i}0^+ )=\frac{1}{N_L}\sum_{\bf q} \psi_d ({\bf k})^2 \frac{b^2 t_{\bf k}^2 t_{{\bf q}-{\bf k}}^2}{((\pi T)^2 +\omega_0^2 )^2} \frac{1}{\pi} \int_{-\infty}^\infty {\rm d}x \nonumber \\
\times [ f(x) X({\bf q},x+\omega +{\rm i}0^+ ) {\rm Im} G ({\bf q}-{\bf k},x+{\rm i}0^+ ) \nonumber \\
 - g(x) {\rm Im} X({\bf q},x+{\rm i}0^+ ) G ({\bf q}-{\bf k}, x-\omega -{\rm i}0^+ )],
\end{eqnarray}
with
\begin{equation}
X({\bf q},\omega +{\rm i}0^+ )=v_d^2 \chi ({\bf q},\omega +{\rm i}0^+)/(1-v_d \chi ({\bf q},\omega +{\rm i}0^+)).
\end{equation}
The renormalized Green's function is given by
\begin{equation}
\widetilde{G} ({\bf k},\omega +{\rm i}0^+ )=[G ({\bf k},\omega +{\rm i}0^+ )^{-1} -\Sigma ({\bf k},\omega +{\rm i}0^+)]^{-1}.
\end{equation}

Now we replace $G$ in eqs. (10) and (11) by $\widetilde{G}$ given in eq. (13).
Then, eqs. (10)-(13) form a set of self-consistent equations for a given pairing interaction $v_d$.
This procedure is called the self-consistent $t$-matrix approximation.
However, it is insufficient for the present case.
The reasons are as follows.
The single-particle spectrum obtained above has a pseudogap, {\it i.e.}, suppression of the density of states near the Fermi energy in the normal state.
The pseudogap suppresses the AF fluctuation and hence the pairing interaction.
Thus, these effects must be treated in a fully consistent manner.
%We expect those effects to be extremely important in high-$T_c$ superconductors because of the enhanced SC fluctuations, especially in the underdoped region.
%Now we evaluate the suppression of the AF fluctuation by the pseudogap self-consistently 
To this end, we take the following scenario.
We calculate $\chi_{\bf s}^{(0)}({\bf q},0)$ in eq. (8) where $G$ is replaced by $\widetilde{G}$ obtained above.
Then, we find the pairing interaction $v_d$ modified by the pseudogap through eqs. (7) and (9).
We repeat the entire process until we obtain a self-consistent solution for $v_d$.
This procedure is called the extended self-consistent $t$-matrix approximation.

Using the solutions obtained above, we determine the superconducting temperature $T_{\rm c}$ by the Thouless criterion,
\begin{equation}
1-v_d \chi ({\bf 0},0) - v_d \chi_{\rm h} ({\bf 0},0) =0,
\end{equation}
where $\chi_{\rm h}$ is due to higher-order terms of the SC fluctuation, which is important in high-$T_c$ superconducting materials because of the quasi 2-dimensionality of ${\rm CuO}_2$ electronic systems.~\cite{SatoS}
In the normal state, close to the superconducting phase, $\chi_{\rm h} ({\bf q},0)$ has a sharp $q$-dependence, practically, $v_d \chi_{\rm h} ({\bf q},0)$ is of $O(1)$ only at ${\bf q} \cong {\bf 0}$ and is zero otherwise in the limit of $T \rightarrow T_{\rm c}$.
Thus, $\chi_{\rm h}$ does not contribute to the ${\bf q}$-summation in the self-energy $\Sigma$ (eq.(11)) in this limit.
In the present study, we simply assume that $v_d \chi_{\rm h} ({\bf 0},0)$ is $0.5$.

We also calculate the NMR relaxation rate, $1/T_1$, using the formula\cite{Imai} 
\begin{eqnarray}
 \frac{1}{T_1 T}=\frac{1}{N_L}\sum_q F_{ab}({\bf q}) \lim_{\omega \rightarrow 0} {\rm Im} \chi^{\rm RPA}_\qb(\omega ) /\omega, \label{T1}
\end{eqnarray}
with
\begin{eqnarray}
\chi^{\rm RPA}_\qb (\omega ) &=& \chi_s^{(0)} (\qb, \omega) /[1-J_{\rm s} ({\bf q}) \chi_s^{(0)} (\qb, \omega)], \label{chiRPA} \\ 
F_{ab}({\bf q}) &=& [A_{ab} +2B (\cos q_xa + \cos q_ya )]^2 /2 \nonumber \\
  &+& [A_c +2B (\cos q_xa + \cos q_ya)]^2 /2, \label{Fab} \\ 
A_{ab} &=& -170, \ \ A_c =20 \ \ {\rm and } \ \ B=40. \label{Aab} 
\end{eqnarray}
Here, $\chi_s^{(0)} (\qb, \omega)$ is given by eq. (8), where $G$ is replaced by $\widetilde{G}$.

In actual numerical calculation, throughout the present study, we set $2t = 1.0$ (which is of $O(1 {\rm eV})$ in real systems), $\Delta = 2.5$ and $\Js = 0.1$. 
The total number of discrete points taken for the $\qb$-summation over the 2D first Brillouin zone is 32 $\times$ 32.
The $\omega$-integral over the region from $-2\omega_0$ to $2\omega_0$ is replaced by the $\omega$-summation of $80$ discrete points.

Numerical results for the $T$-dependence in $1/T_1T$ are shown in Fig. 1 for the hole-doping rate $\delta=0.06 \sim 0.19$.
At high temperatures, $1/T_1T$ increases as $T$ or $\delta$ decreases, due to the development of the AF fluctuation.
We define $T_{\rm 0}$ as the characteristic temperature where the effects of the AF fluctuation begin to be appreciable.
Explicit values of $T_{\rm 0}$ are determined by the condition that $[\frac{\partial}{\partial T} (1/T_1 T) \vert_{T=T_{\rm 0}}]/[(1/T_1 T) \vert_{T\rightarrow \infty, \delta =0.1}] =-2 \times 10^2$.
At $\delta < 0.11$, $1/T_1T$ has a maximum at a certain temperature denoted by $T_{\rm sg}$.
Curves of $1/T_1T$ on the lower temperature sides terminate at $T_{\rm c}$ for $\delta \simg 0.1$, but we have not yet obtained $T_{\rm c}$ for $\delta \siml 0.09$ due to the poor convergency of the numerical calculation at $T\cong T_{\rm c}$.
In the inset, the $T$-dependence of $1/T_1T$ obtained at $\delta =0.1$ as above (solid line) is compared with that calculated by neglecting the effects of the SC fluctuation (dashed line).
We note that the SC fluctuation also becomes appreciable at $T \sim T_{\rm 0}$ and continues to develop along with the AF fluctuation as $T$ decreases, while the SC fluctuation induces the pseudogap in the single-particle spectra (see Figs. 3 and 4), which suppresses the AF fluctuation.
Eventually, the developed pseudogap eliminates the AF fluctuation, which leads to a sharp drop in $1/T_1T$ at $T \siml T_{\rm sg}$.
Recently, Kontani showed that the Hall coefficient $R_{\rm H}$ is also enhanced by the AF fluctuation.\cite{Kontani}
Thus, the initial enhancement of $R_{\rm H}$ at high temperatures can be assumed to have the same origin as that of $1/T_1 T$.
In this sense, we may argue that $T_{\rm 0}$ obtained in the present study corresponds to $T_{\rm 0}$ determined by $R_{\rm H}$ experiments.~\cite{Nishikawa}

\begin{figure}
\def\epsfsize#1#2{0.4#1}
\centerline{
\epsfbox{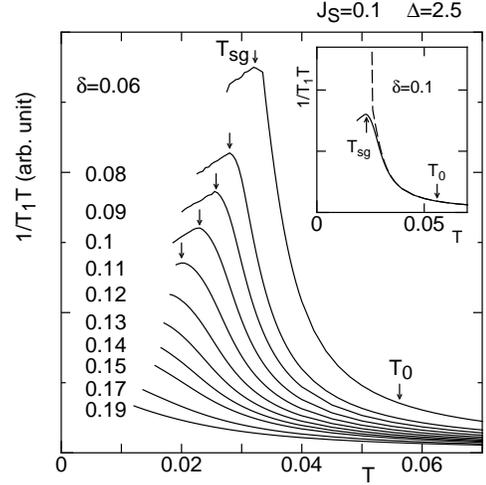}
%\hspace*{0.3cm}
}
\caption{
$T$-dependences of $1/T_1 T$ obtained with $J_{\rm s} = 0.1$ and $\Delta =2.5$ for the hole-doping rate $\delta =0.06 \sim 0.19$.
In the inset, $1/T_1T$ (solid line) is compared with that calculated by neglecting the effects of the SC fluctuation (dashed line), where $\delta =0.1$.
}
\label{fig.1}
\end{figure}

%We have also calculated the knight shift $K \propto \chi_{{\bf q}={\bf 0}}^{\rm RPA} (\omega =0)$.
%The knight shift $K$ corresponds to the average of the density of states near the Fermi energy in a width which is about $T$.
%As temperature decreases, $K$ increases slightly because of singular shape of density of states in the 2D system.
%At the region of $\delta \siml 0.13$ $K$ turn to decrease in low temperature region.
%It indicates that decrease of the density of states near the Fermi energy by SC fluctuation, {\it i.e.} a pseudo (spin) gap effect.
% $T_{\rm sg^\prime}$ は$T_{\rm sg}$よりわずかに高い。
%しかし実験ではむしろ$T_{\rm sg^\prime}$は$T_{\rm sg}$に比べてずっと高温にある。
%オーダーとしては$T_{\rm sp}$に近い。
% $T_{\rm sg^\prime}$ 、$K$が最大になる温度、は計算上あいまいなところがある。
%なぜなら$T_{\rm sg^\prime}$ は揺らぎを入れない時の$K$の温度依存性に大きく依存するからである。
%揺らぎを入れない時の$K$の温度依存性は準粒子バンドの状態密度のフェルミ面近傍の形状に依存する。
%本研究の準粒子バンドは最近接d-p混成のみの2次元ｄ−ｐ模型なので、状態密度の形状は物質に対応しない。

We plot the three characteristic temperatures $T_{\rm 0}$, $T_{\rm sg}$ and $T_{\rm c}$ as functions of $\delta$ in Fig. 2.
Note that $T_{\rm sg}$ appears at $\delta \cong 0.11$ and increases as $\delta$ decreases, whereas $T_{\rm 0}$ appears at $\delta \cong 0.26$ and is much higher than $T_{\rm sg}$.
We also note that $T_{\rm c}$ has a maximum at $\delta \cong 0.11$.
The dotted curve of $T_{\rm c}$ at $\delta \siml 0.09$ is determined by eq. (14) with the help of the extrapolation of numerical results for the $T$-dependence of ($1-v_d \chi ({\bf 0},0) - v_d \chi_{\rm h} ({\bf 0},0)$), because we had difficulty in obtaining fully convergent solutions for the self-consistent equations (7)-(13) at $T \cong T_{\rm c}$ for $\delta \siml 0.09$.
As $\delta$ decreases, the quasi-particle band narrows and the nesting effect increases, which enhances the AF fluctuation and then pushes $T_{\rm 0}$ higher.
The AF fluctuation enhances the pairing interaction and then the SC fluctuation, which develops the pseudogap and then suppresses $T_{\rm c}$ in the underdoped region.
$T_{\rm sg}$ is determined by the competition between the AF fluctuation and the SC fluctuation.
The phase diagram shown in Fig. 2 accounts for essential features of the anomalous metallic phase observed in high-$T_{\rm c}$ superconducting materials.~\cite{Sato}
%The key factors in obtaining the reasonable features of $\delta$-dependence of $T_{\rm c}$ and $T_{\rm sg}$ is due to the suppression of the AF fluctuation by the pseudogap and due to the use of the quasi-particles with the band width proportional to $\delta$ at the underdoped region.

\begin{figure}
\def\epsfsize#1#2{0.4#1}
\centerline{
\epsfbox{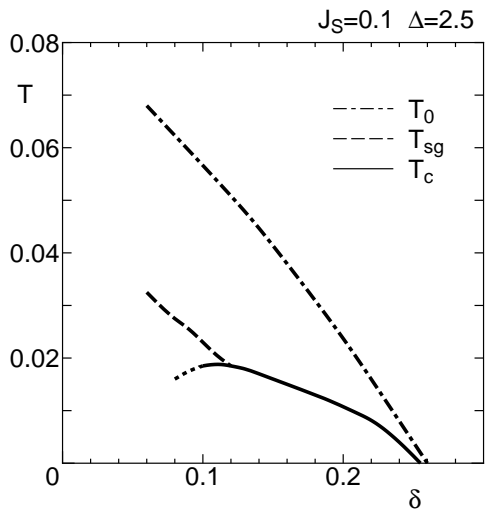}
%\hspace*{0.3cm}
}
\caption{
$\delta$-$T$ phase diagram with $J_{\rm s} = 0.1$ and $\Delta =2.5$.
}
\label{fig.2}
\end{figure}

The single-particle density of states, $\rho (\omega ) = -\frac{1}{\pi}{\rm Im} \frac{1}{N_{\rm L}} \sum_{\bf k} \widetilde{G} ({\bf k},\omega +{\rm i}0^+ )$, at $T=0.02$, $0.025$, $0.03$ and $0.2$ with $\delta =0.1$ where $T_{\rm 0} =0.057$ are shown in Fig. 3.
At $T=0.2$, $\rho (\omega )$ is nearly equal to that of the bare in-gap states without the effects of the SC fluctuation.
At $T \siml T_{\rm 0}$, $\rho (\omega )$ decreases near the Fermi energy ($\omega =0$) and increases at $\vert \omega \vert \simg 0.05$ as $T$ decreases.
The size of the pseudogap is roughly $0.1$, which corresponds to the strength of the pairing interaction of order of $J_{\rm s}$.
%強いSingularutyのあるDOSのためはっきりしたPseudogapが出ないが、Non-interactingとの比で見ればはっきりとPseudogapになっている。

\begin{figure}
\def\epsfsize#1#2{0.4#1}
\centerline{
\epsfbox{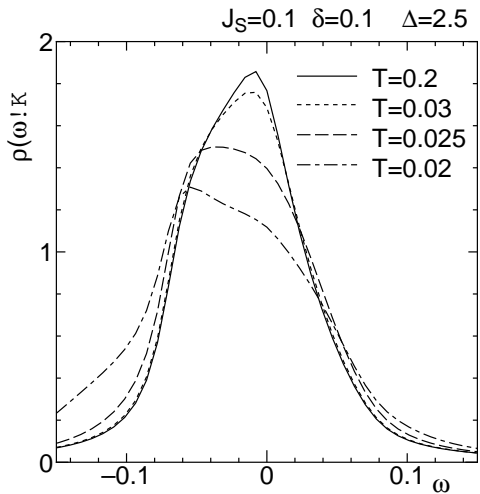}
%\hspace*{0.3cm}
}
\caption{
Single-particle density of states at $T=0.02$, $0.025$, $0.03$ and $0.2$ with $\delta =0.1$, $J_{\rm s} = 0.1$ and $\Delta =2.5$, where $T_{\rm 0}=0.057$.
}
\label{fig.3}
\end{figure}

Lastly, we show contour plots of the ${\bf k}$-dependence on the first Brillouin zone of the single-particle spectrum at the Fermi energy $\rho ({\bf k},0)=-\frac{1}{\pi} {\rm Im} G({\bf k},0)$  with $\delta =0.1$ at $T=0.2$ in Fig. 4 (left) and at $T=0.02$ in Fig. 4 (right).
The single-particle spectrum $\rho ({\bf k},0)$ has a large value in bright regions, where the energy of a quasi particle with ${\bf k}$ is close or equal to the Fermi energy.
Thus, the bright regions represent the shape of the Fermi surface.
In the dark regions, $\rho ({\bf k},0)$ is nearly zero, where the energy of a quasi particle with ${\bf k}$ is far from the Fermi energy.
The Fermi surface exists in all directions at $T=0.2$.
As the temperature decreases, the single-particle spectra near the $[\pi ,0]$ and $[0,\pi ]$ directions decrease due to the SC fluctuation with $d_{x^2 -y^2}$ symmetry, and then the Fermi surface remains only near the $[\pi ,\pi ]$ direction at $T=0.02$.
These features are consistent with the results of ARPES experiments.\cite{Norman}

The present work has been supported by a Grant-in-Aid for Scientific Research from the Ministry of Education, Science, Sports and Culture.

\begin{figure}
\def\epsfsize#1#2{0.4#1}
\centerline{
\epsfbox{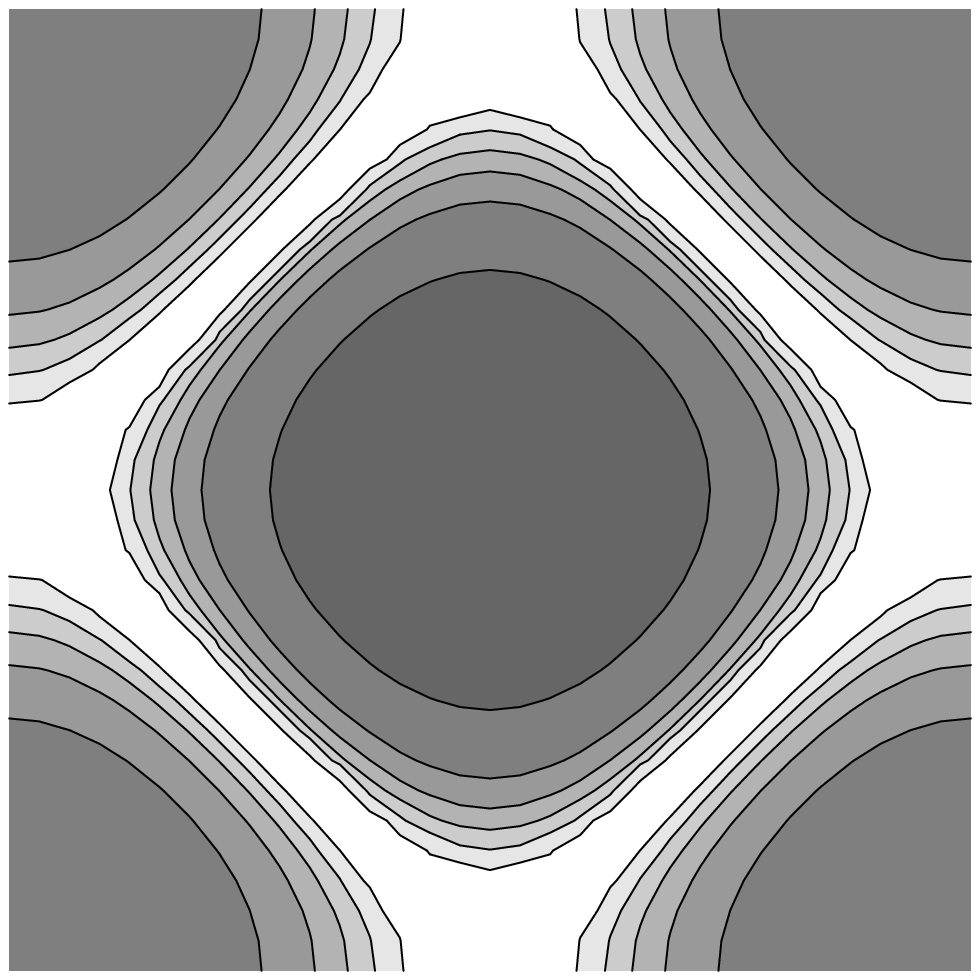}
\epsfbox{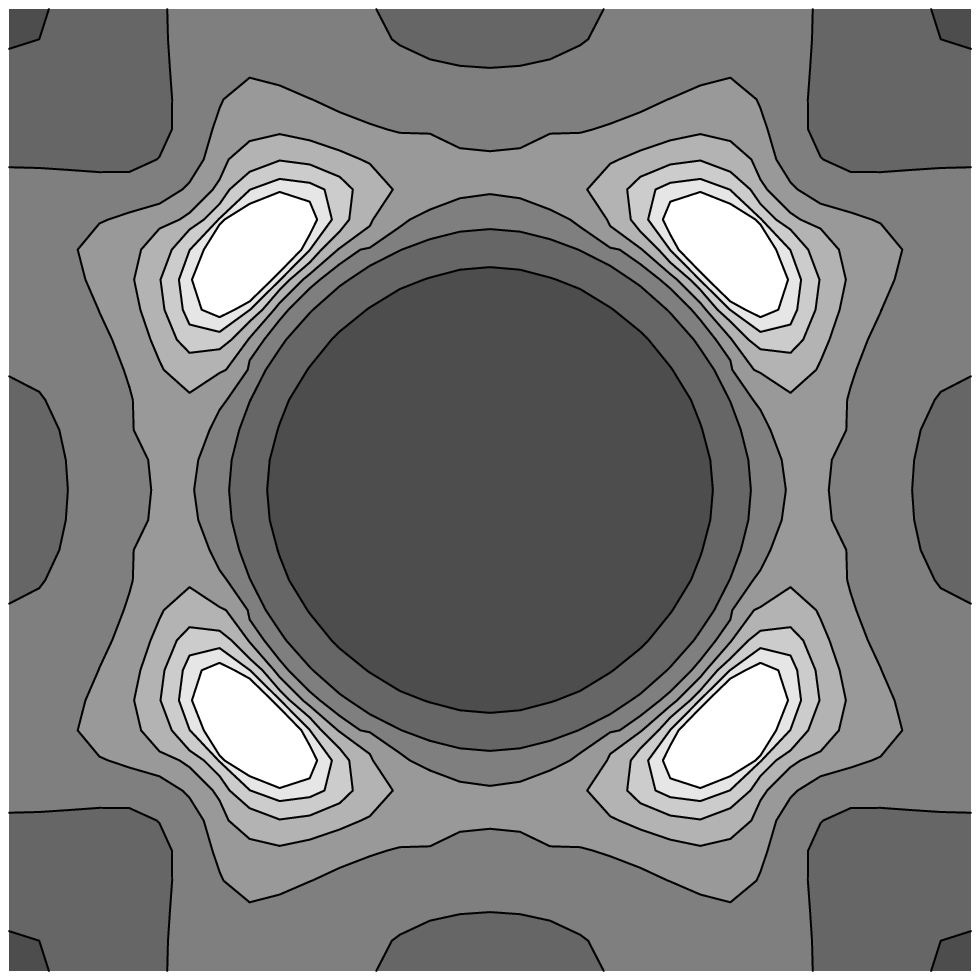}
%	\hspace*{0.8cm}
}
\caption{
Contour plots of ${\bf k}$-dependences (first Brillouin zone) of the single-particle spectra at Fermi energy $\rho ({\bf k}, \omega =0)$ with $\delta =0.1$, $J_{\rm s} = 0.1$ and $\Delta =2.5$, where $T=0.2$ (left) and $T=0.02$ (right).
Bright regions represent the shape of the Fermi surface.
}
\label{fig.4}
\end{figure}

\end{document}